\documentclass[aps,superscriptaddress,twocolumn,prb,preprintnumbers,amsmath,amssymb,nofootinbib]{revtex4-1}
\usepackage{graphicx,afterpage}
\usepackage{multirow}
\usepackage{bm}
\usepackage{xcolor}
\usepackage[T1]{fontenc}

\newcommand{\be}{\begin{eqnarray}}
\newcommand{\ee}{\end{eqnarray}}

\newcommand{\vs}[1]{ \boldsymbol{#1} }

\begin{document}

\title{Singular magnetic anisotropy in the nematic phase of FeSe}

%-------------------------------------------------------------------------------------
\author{Rui Zhou}
\email{rzhou@iphy.ac.cn}
\affiliation{Univ. Grenoble Alpes, INSA Toulouse, Univ. Toulouse Paul Sabatier, EMFL, CNRS, LNCMI, 38000 Grenoble, France}
%-------------------------------------------------------------------------------------

%-------------------------------------------------------------------------------------
\author{Daniel D. Scherer}
\affiliation{Niels Bohr Institute, University of Copenhagen, Lyngbyvej 2, 2100 Copenhagen, Denmark}
%-------------------------------------------------------------------------------------

%-------------------------------------------------------------------------------------
\author{Hadrien Mayaffre}
\affiliation{Univ. Grenoble Alpes, INSA Toulouse, Univ. Toulouse Paul Sabatier, EMFL, CNRS, LNCMI, 38000 Grenoble, France}
%-------------------------------------------------------------------------------------

%-------------------------------------------------------------------------------------
\author{Pierre Toulemonde}
\affiliation{Universit\'e Grenoble Alpes, CNRS, Grenoble INP, Institut N\'eel, F-38000 Grenoble, France}
%-------------------------------------------------------------------------------------

%-------------------------------------------------------------------------------------
\author{Mingwei Ma}
\affiliation{International Center for Quantum Materials, School of Physics, Peking University, Beijing 100871, China}
%-------------------------------------------------------------------------------------

%-------------------------------------------------------------------------------------
\author{Yuan Li}
\affiliation{International Center for Quantum Materials, School of Physics, Peking University, Beijing 100871, China}
\affiliation{Collaborative Innovation Center of Quantum Matter, Beijing 100871, China}
%-------------------------------------------------------------------------------------

%-------------------------------------------------------------------------------------
\author{Brian M. Andersen}
\email{bma@nbi.ku.dk}
\affiliation{Niels Bohr Institute, University of Copenhagen, Lyngbyvej 2, 2100 Copenhagen, Denmark}
%-------------------------------------------------------------------------------------

%-------------------------------------------------------------------------------------
\author{Marc-Henri Julien}
\email{marc-henri.julien@lncmi.cnrs.fr}
\affiliation{Univ. Grenoble Alpes, INSA Toulouse, Univ. Toulouse Paul Sabatier, EMFL, CNRS, LNCMI, 38000 Grenoble, France}
%-------------------------------------------------------------------------------------

\date{\today}

\begin{abstract}

FeSe is arguably the simplest, yet the most enigmatic, iron-based superconductor. Its nematic but non-magnetic ground state is unprecedented in this class of materials and stands out as a current puzzle. Here, our NMR measurements in the nematic state of mechanically detwinned FeSe reveal that both the Knight shift and the spin-lattice relaxation rate $1/T_1$ possess an in-plane anisotropy opposite to that of the iron pnictides LaFeAsO and BaFe$_2$As$_2$. Using a microscopic electron model that includes spin-orbit coupling, our calculations show that an opposite quasiparticle weight ratio between the $d_{xz}$ and $d_{yz}$ orbitals leads to an opposite anisotropy of the orbital magnetic susceptibility, which explains our Knight shift results. We attribute this property to a different nature of nematic order in the two compounds, predominantly bond-type in FeSe and onsite ferro-orbital in pnictides. The $T_1$ anisotropy is found to be inconsistent with existing neutron scattering data in FeSe, showing that the spin fluctuation spectrum reveals surprises at low energy, possibly from fluctuations that do not break $C_4$ symmetry. Therefore, our results reveal that important information is hidden in these anisotropies and they place stringent constraints on the low-energy spin correlations as well as on the nature of nematicity in FeSe.

\end{abstract}

\maketitle

%-------------------------------------------------------------------------------------
 \subsubsection{Introduction}
%-------------------------------------------------------------------------------------

How and why electronic degrees of freedom break the point-group symmetry of the crystal lattice is currently one of the most active areas of research in condensed-matter physics~\cite{ARCMP}. This phenomenon fascinates not only as an electronic analog of the nematic phase of liquid crystals (a phase with broken rotational, but not translational, symmetry of the molecular arrangement)~\cite{SNature98} but also because it might help understand several classes of unconventional superconductors~\cite{Yamase13,Maier14,Lederer15,KNature15}.

In recent years, intense research activities have focused on the iron-pnictide superconductors, in which the electronic fluid is manifestly responsible for a tetragonal-to-orthorhombic transition~\cite{Chu2010,Fernandes2014}. Concomitantly, a magnetic transition is also systematically induced at a temperature equal to, or slightly lower than, the structural (now dubbed nematic) transition at $T_s$~\cite{Dai2015}. Even though a microscopic model of magnetism is still debated~\cite{Bascones2016}, the entanglement between magnetic and structural degrees of freedom is largely considered to be the cornerstone of the physics of the pnictides.

It thus came as a surprise that another iron-based superconductor, the iron chalcogenide FeSe ($T_c=9$~K), exhibits an orthorhombic transition that is neither preceded nor followed by any magnetic transition~\cite{Cava2009}. This absence of magnetic order has led to the suggestion that nematic order in FeSe is driven by orbital, rather than spin, fluctuations~\cite{Boehmer2015,Baek2015,Yamakawa2016}. This issue, however, has remained highly controversial. First, despite the absence of magnetic order, strong spin fluctuations are present~\cite{Rahn2015,ZhaoMat2016,Mingwei2016,Zhao2016,Tong2019} and these could drive nematic order~\cite{Schmalian2015}. Further, the magnetic properties of FeSe, including the absence of spin order, are not well understood and defining the correct theoretical model is even more contentious than for the pnictides~\cite{FeSe1,FeSe2,FeSe3,FeSe4,FeSe5,FeSe6,FeSe7,Mukherjee2015,Schmalian2015,Lee2015,Glasbrenner2015,Yamakawa2016,scherer17,Fanfarillo2018,Kreisel2018}.  Last, experiments able to discriminate the different theories are scarce. 

Besides magnetism, the electronic properties of FeSe also appear more mysterious than those of the pnictides. The band structure, with tiny Fermi surface pockets, is strongly renormalized, as compared to DFT calculations~\cite{AnneAndreas2017,MattAmalia2018}. In addition, measurements of low-energy quasiparticle interference in the normal state and of the superconducting gap structure~\cite{Kostin2018,Sprau2017} have shown that the nematic anisotropy is much larger than that prescribed by simple models that include realistic band splitting caused by the nematic order. Several theoretical models have invoked effects from electron interactions in order to capture the observed momentum (or real-space) anisotropy~\cite{Yamakawa2016,Onari16,Nica17,kreisel17,Benfatto18,Kang2018,She17}. At present, however, the detailed low-energy electronic bands for FeSe are still controversial~\cite{Fanfarillo2016,Watson17,MYi,Huh,Rhodes20} and recent studies have pointed to unusual orbital contributions to the nematic order in this material~\cite{XLong,MHC_trans}.

Here, we use nuclear magnetic resonance (NMR) to discover that the in-plane magnetic anisotropy of both the Knight shift $K$ and the spin-lattice relaxation rate $1/T_1$ in the nematic state of FeSe is inverse to that in nematic LaFeAsO.

In LaFeAsO, both the $K$ anisotropy and the $T_1$ anisotropy measured here are consistent with other measurements in pnictides and they are understood from the well-established sequence of events upon cooling: 1) the high temperature tetragonal phase features degenerate spin fluctuations at two orthogonal wave vectors $\vs{Q}_1=(\pi,0)$ and $\vs{Q}_2=(0,\pi)$. 2) The orthorhombic distortion at $T_s$ and the concomitant orbital ordering of Fe $d_{xz}$ and $d_{yz}$ orbitals lift this degeneracy by selecting $\vs{Q}_1$~\cite{Lu2014}. 3) This induces stripe-type magnetic order~\cite{Dai2015} where the relative orientation of magnetic moments becomes effectively ferromagnetic (FM) along the short in-plane axis $b$ and antiferromagnetic (AFM) along the long axis $a$ while SOC enforces the ordered moments to be aligned with the AFM direction~\cite{Dai2015,Christensen2015,scherer18}. Upon cooling, Knight shift anisotropy appears at $T_s$~\cite{Zhou2016}, as does $T_1$ anisotropy~\cite{Fu2012,Kissikov2017}. The former arises from the slight difference of uniform spin susceptibility $\chi_{\rm spin}(\vs{q}=\vs{0})$ along the $a$ and $b$ axes~\cite{He2016}. The latter is different in nature and results from two combined effects: the imbalance of spectral weight of spin fluctuations at $\vs{Q}_1$ and $\vs{Q}_2$ and the anisotropy of the spatial components of the spin (essentially $\chi_c^{\prime\prime}(\vs{Q}_1)\neq \chi_{ab}^{\prime\prime}(\vs{Q}_1)$)~\cite{Kissikov2017}.

The contrasting results in FeSe imply significant deviations from this picture. From detailed theoretical modeling using realistic band structures and including SOC and the feedback of the nematic order on the self-energy, we find that a consistent description of the Knight shift anisotropy at $\vs{q}=(0,0)$ can be obtained by invoking different nematic orders in LaFeAsO and FeSe. This highlights the important information hidden in the magnetic anisotropy. The $T_1$ anisotropy, on the other hand, finds no straightforward explanation within a phenomenological model that considers assumed functional forms of the main contributors to the magnetic susceptibility.

The paper is organized as follows: in Sec.~\ref{Sec:principle} we outline the principle of the experiment, while Sec.~\ref{Sec:K-results} and Sec.~\ref{Sec:T1-results} contain the experimental results for the in-plane anisotropy of the Knight shift and the relaxation rate, respectively. In Sec.~\ref{Sec:theoryKS} and Sec.~\ref{Sec:theoryT1} we provide theoretical analyses of the Knight shift and the relaxation rate, respectively. The Knight shift is calculated in terms of a microscopic approach whereas the relaxation rate is discussed from a more phenomenological point of view. Section \ref{Sec:discussion} includes a discussion of the results and connection to the literature. We note that there exists a substantial associated supplementary material (SM) section with additional details of the experimental setup and the theoretical analyses.

%
%-------------------------------------------------------------------------------------
%-------------------------------------------------------------------------------------
%

%\section{Results}

%-------------------------------------------------------------------------------------
\subsubsection{Principle of the experiment}
\label{Sec:principle}
%-------------------------------------------------------------------------------------

The principle of our measurements is simple: the presence of perpendicular domains in the orthorhombic phase hampers an unambiguous interpretation of the NMR spectra. Therefore, we mechanically detwinned the crystals by applying uniaxial stress (see Methods). The short axis, labeled $b$ by convention, is determined here by the direction of applied strain. This allows us to determine which NMR lines correspond to sites having the magnetic field $H$ aligned with the $a$ or $b$ axis. This unambiguous site assignment is then used to evaluate both the amplitude and the sign of the anisotropy of two NMR observables: the Knight shift $K$ and the spin-lattice relaxation rate $1/T_1$. Note, however, that these anisotropies have been measured directly in the twinned samples and so are not caused by uniaxial stress.

In the absence of strong spin-orbit coupling (SOC), the Knight shift tensor $K$ is usually decomposed into a spin and an orbital contribution :
\begin{equation}
\label{f1}
\begin{aligned}
       K_{\alpha \alpha }= K_{\alpha \alpha }^{\rm spin}+K_{\alpha \alpha }^{\rm orb}
\end{aligned}
\end{equation}
The spin contribution $K^{\rm spin}$ is proportional to the real part of the static ($\omega=0$), uniform ($q=0$) spin susceptibility $\chi^{\rm spin}$ and to the hyperfine coupling tensor $A_{\alpha \alpha}$:
\begin{equation}
\label{f1b}
\begin{aligned}
K_{\alpha \alpha }^{\rm spin} \propto A_{\alpha \alpha }^{\text{spin}}\chi_{\alpha\alpha }^{\rm spin},
\end{aligned}
\end{equation}
while $K^{\rm orb}$ includes effects of the orbital susceptibility $\chi^{\rm orb}$ due to the orbital momentum $L$ and closed-shell diamagnetism and (again in the absence of SOC), the total uniform magnetic susceptibility is:
\begin{equation}
\label{f1c}
\begin{aligned}
       \chi^{\rm mag}= \chi^{\rm spin}+\chi^{\rm orb}
\end{aligned}
\end{equation}

The spin-lattice relaxation rate $1/T_1$ is usually dominated by low-energy spin fluctuations and is related to the imaginary part of the dynamic spin susceptibility ${\chi}^{\prime\prime}\left( \vs{q},{{\omega }_{0}} \right)$:
\begin{equation}
\label{fs1}
\frac{1}{T_1}=T\frac{{{\gamma }^{2}}}{2}\sum\limits_{\vs{q}}{\sum\limits_{\alpha }{{{\mathcal{F}}_{\alpha \alpha }}\left( \vs{q} \right)\frac{{{\chi}^{\prime \prime }_{\alpha\alpha}}\left( \vs{q},{{\omega }_{0}} \right)}{{{\omega }_{0}}}}},
\end{equation}
where $\gamma$ is the nucleus gyromagnetic ratio, $\omega_0$ the nuclear Larmor frequency ($\sim10^8$~Hz here) and ${{\mathcal{F}}_{\alpha \alpha }}\left( \vs{q} \right)={{\left| {{A}_{\alpha \alpha }}\left( \vs{q} \right) \right|}^{2}}$ the hyperfine form factor ($\alpha=a,b,c$) with ${{A}_{\alpha \alpha }}\left( \vs{q} \right)=\sum\limits_{j}{{{A}_{\alpha \alpha }}\left( {{{\vs{r}}}_{j}} \right)}\exp \left( -i\vs{q}\cdot {{{\vs{r}}}_{j}} \right)$, $j$ labels the four Fe around each As/Se site.

Since large single crystals of FeSe cannot be grown, neutron scattering data from detwinned samples is limited at present\cite{Tong2019}. Therefore, NMR measurements under uniaxial stress offer a unique opportunity to probe the anisotropy of spin fluctuations in the mysterious nematic state. In order to provide a benchmark for our $^{77}$Se NMR results in FeSe, we  systematically compare them with $^{75}$As data in the parent compound LaFeAsO. We have chosen this latter compound because it offers a relatively wide temperature ($T$) window between its orthorhombic transition at $T_s\simeq 154$~K ($T_s\simeq 90$~K for FeSe) and its magnetic transition at $T_N\simeq135$~K~\cite{Cruz2008}. We show that data from different probes of magnetism in LaFeAsO or BaFe$_2$As$_2$ agree quantitatively with our NMR results in LaFeAsO, implying that these results indeed represent the typical behavior of iron-pnictides~\cite{Dai2015}.

\begin{figure*}[t!]
%\vspace{3mm}
\centerline{\includegraphics[width=6in]{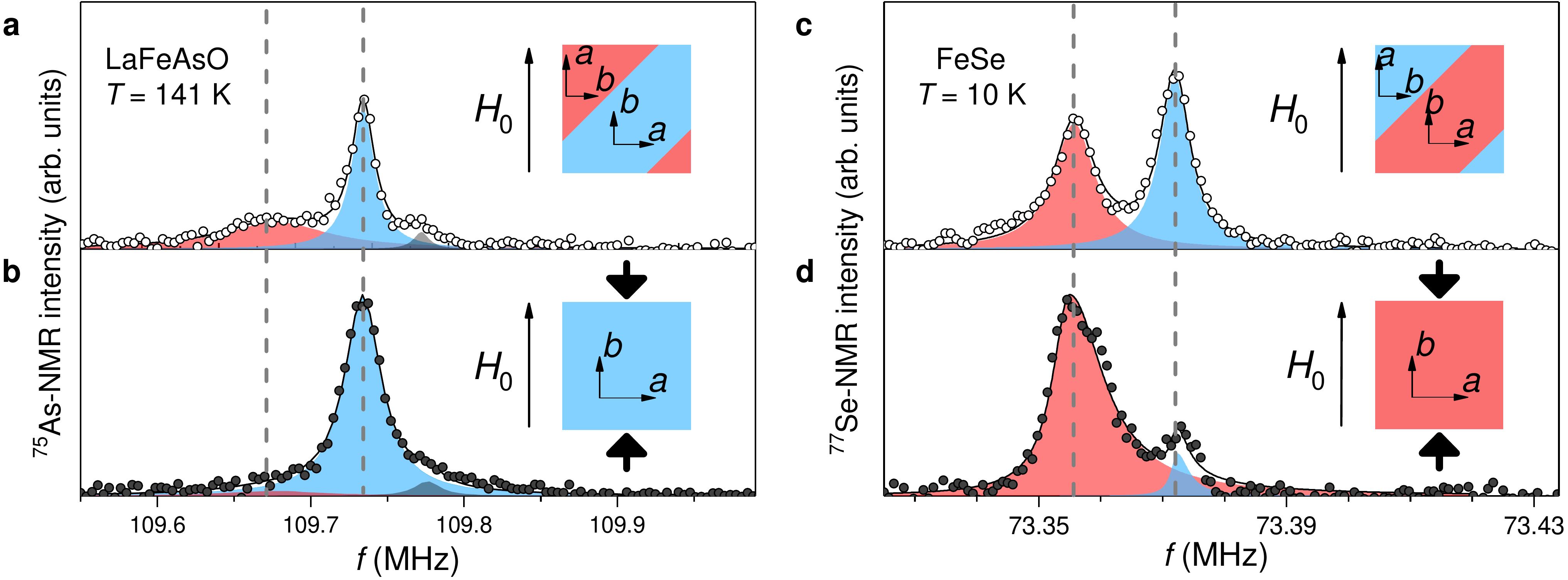}} %%%%%%%%%%%%%%%
\caption{{\bf NMR evidence for sample detwinning under uniaxial stress.} (a,b) $^{75}$As-NMR central line in LaFeAsO with (b) and without (a) uniaxial stress at $T = 141$~K and $H_0 = 15$~T. Upon application of stress, the integrated intensity of the higher-frequency peak increases at the expense of the low frequency peak. (c,d) $^{77}$Se-NMR spectra of FeSe  with (d) and without (c) stress at $T$ = 10 K and $H_0$ = 9 T, respectively. At variance with LaFeAsO, the remaining intensity of the second peak in FeSe indicates that the crystal is not fully detwinned. This is because the maximum stress applicable on FeSe crystals is about half of that for LaFeAsO. FeSe distorts and exfoliates above $\sim$ 5~MPa while LaFeAsO does not break up to at least 15~MPa. This thus limits the maximal stress that can be applied on FeSe that we could not fully detwin it. LaFeAsO spectra are fitted by three Lorentz functions (the small peak at $\sim$109.77~MHz is present at any temperature and is attributed to an impurity - see SM). FeSe spectra are fitted by two Lorentz functions. The 1:1 area ratio of the peaks in (a) and (c) shows that the twin domains are equally populated in the unstrained samples. The different widths for the two peaks is consistent with other reports (e.g.ref. ~\cite{Ok2018}) but not presently understood. Notice that the quadrupolar interaction contributes to the peak splitting in LaFeAsO. As explained in SM, this effect has been taken into account for extracting the Knight shift anisotropy in Fig.~\ref{asy}. }
\label{unispectrum}
\end{figure*}

%-------------------------------------------------------------------------------------
\subsubsection{Knight shift anisotropy: experimental results}
\label{Sec:K-results}
%-------------------------------------------------------------------------------------

When no stress is applied and the field is aligned with the Fe-Fe bonds ({\it i.e.} the orthorhombic $a$ or $b$ axis), NMR lines in the (non-magnetic) orthorhombic phase are split into two peaks (Fig.~\ref{unispectrum}a,c), in agreement with earlier studies of both LaFeAsO and FeSe~\cite{Fu2012,Baek2015,Boehmer2015,Ok2018,rxcao2018,Tao2020}. Upon applying uniaxial stress in both samples, the intensity of one peak grows at the expense of the other while the total intensity of the spectrum is conserved, as seen from Fig.~\ref{unispectrum}(b,d). The second peak is strongly reduced at 5~MPa (the maximum pressure that FeSe can withstand) and eventually disappears above $\sim$10-20~MPa in LaFeAsO. This demonstrates that the line splitting in the unstrained crystals is produced by twin domains, as previously hypothesized~\cite{Baek2015,Fu2012,Kitagawa2008}: one peak corresponds to those domains for which the magnetic field $H$ is parallel to the $a$-axis and the other peak corresponds to domains with $H\parallel b$. That each domain is associated with a single NMR peak shows that the orbital order does not lead to any differentiation of Se sites in each domain (at least when the field is applied parallel to the $a$ or $b$ axes).

For FeSe, the resonance frequency is entirely determined by the Knight shift $K$ so the difference of resonance frequencies for $H\parallel a$ and $H\parallel b$ is due to the anisotropy of $K$ in the plane. Because $^{77}K > 0$, the low-frequency peak corresponds to the lowest $K$ value. For LaFeAsO, after deconvolution from quadrupole effects that are present due to the nuclear spin $I=3/2$ of $^{75}$As, our analysis (see SM) shows that significant Knight shift anisotropy also contributes to the line splitting. The low-frequency peak corresponds to the lowest $K$ values, as in FeSe.

\begin{figure*}[t!]
%\vspace{3mm}
\centerline{\includegraphics[width=6in]{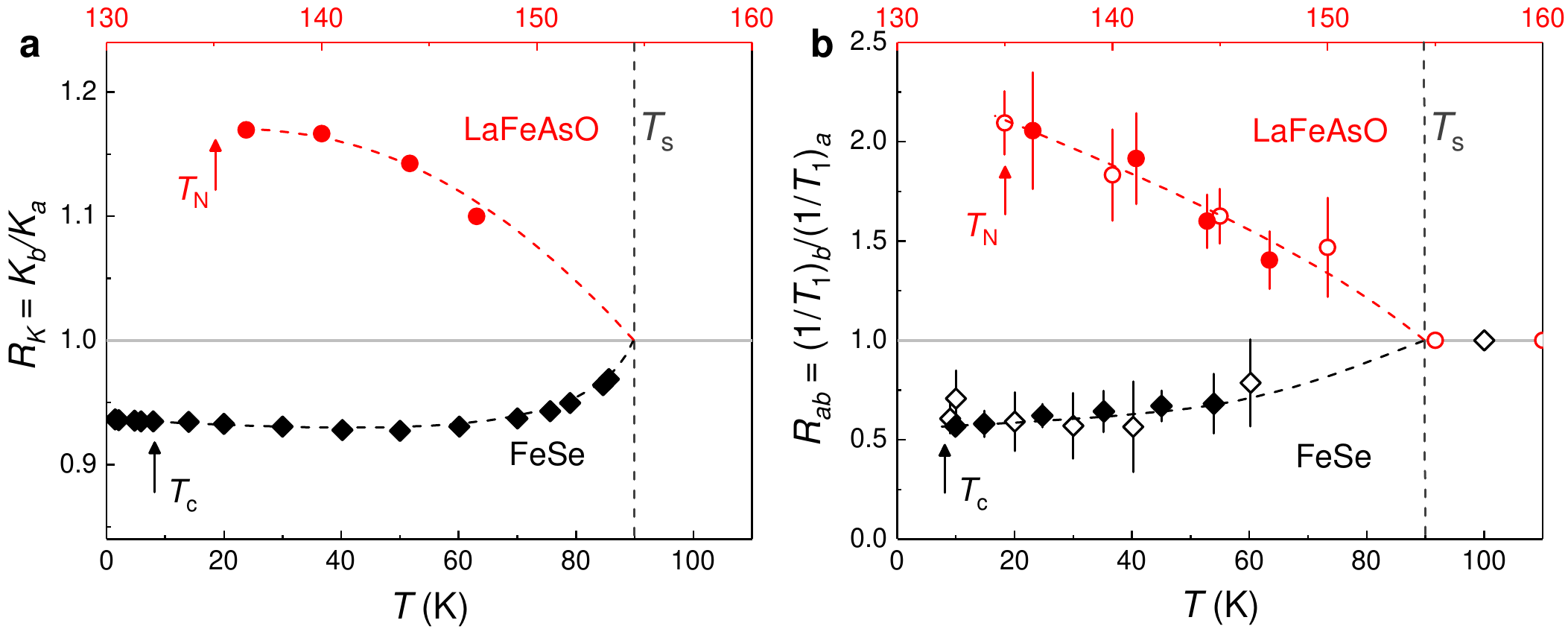}} %%%%%%%%%%%%%%%
\caption{{\bf Knight shift and $T_1$ anisotropies in the nematic state.} Temperature dependence of in-plane Knight shift anisotropy ratio $K_b/K_a$ (a) and in-plane 1/$T_1$ anisotropy $R_{ab} = (1/T_1)_b/(1/T_1)_a$ (b). Our data (solid symbols) are fully consistent with Refs.~\onlinecite{Fu2012,Baek2015} (open symbols) but because we detwinned the crystals, our data are able to unambiguously identify the $a$ and $b$ axes. Dotted lines are guides to the eye. Error bars in (b) correspond to s.d. in fits of the recovery of the nuclear magnetization after a saturating pulse.}
\label{asy}
\end{figure*}

Despite these qualitative similarities, however, there is a remarkable difference between the pnictide- and the chalcogenide-based materials: while applied stress parallel to the external field direction forces all sites to have $H\parallel b$ (which is the short axis by definition) in both compounds, the peak that remains in the detwinned crystal is the high frequency peak in LaFeAsO while it is the low frequency peak in FeSe as shown in Fig.~\ref{unispectrum}. Therefore, both compounds develop in-plane Knight shift anisotropy on cooling below $T_s$ (see Fig.~\ref{asy}(a) where we plot the ratio $R_K=K_b/K_a$), but their anisotropies are inverted: $K_a>K_b$ for FeSe and $K_b>K_a$ for LaFeAsO (from now on we use $K_{\alpha\alpha}=K_\alpha$).

%-------------------------------------------------------------------------------------
%-------------------------------------------------------------------------------------

In principle, both the hyperfine coupling and the magnetic susceptibility are susceptible to become anisotropic in response to the orthorhombic distortion and to the orbital imbalance below $T_s$. However,  there are indications from BaFe$_2$As$_2$ that the anisotropy of $K$ arises mostly from the anisotropy of $\chi^{\mathrm{mag}}$: indeed, comparing NMR~\cite{Kissikov2017} and bulk magnetic susceptibility data~\cite{He2016} at $\sim$138~K, we find that $K_b/K_a$ and $\chi^{\mathrm{mag}}_b/\chi^{\mathrm{mag}}_a$ change at the same rate as a function of relative lattice distortion. Furthermore, we observe that $K_b/K_a\simeq 1.17$ at $T\simeq T_N$ in LaFeAsO has the same value as in BaFe$_2$As$_2$~\cite{Kissikov2017} and NaFeAs~\cite{Zhou2016}. Therefore, all iron-pnictides appear to have the same value of the Knight shift anisotropy at $T_N$ and this anisotropy is essentially due to the anisotropy of $\chi^{\mathrm{mag}}$. Thus, we can now assess that
\begin{equation}
\label{expanisotropyLaFeAsO}
\chi^{\mathrm{mag}}_b > \chi^{\mathrm{mag}}_a~\text{in LaFeAsO},
\end{equation}
that is to say, the uniform susceptibility is larger for $H\parallel b$ than for $H\parallel a$, as also found in BaFe$_2$As$_2$~\cite{He2016}.

Since $^{77}$Se nuclei in FeSe and $^{75}$As in LaFeAsO occupy the same crystallographic position with respect to the Fe square lattice, their hyperfine coupling should have the same symmetry properties. Therefore, the Knight shift anisotropy in FeSe must also arise from the anisotropy of $\chi^{\mathrm{mag}}$. This leads us to conclude that the anisotropy of the uniform spin susceptibility in FeSe is reversed with respect to that in iron pnictides:
\begin{equation}
\chi^{\mathrm{mag}}_a > \chi^{\mathrm{mag}}_b~\text{in FeSe}.
\label{chi_FeSe}
\end{equation}

This conclusion corroborates a recent bulk measurement~\cite{Mingquan2017} finding a ratio $\chi^{\mathrm{mag}}_b / \chi^{\mathrm{mag}}_a=0.93$. The NMR shift being insensitive to extrinsic contributions such as diluted impurities or spurious phases, the intrinsic nature of the effect is not questionable. Furthermore, that our $K_b/K_a$ ratio at low temperature gives the same value within error bars (Fig.~\ref{asy}a) strongly suggests that the magnetic anisotropy arises entirely from either the spin part or the orbital part, rather than from a combination of the two. Indeed, the relative weight of these two parts in the total Knight shift (Eqs.~\ref{f1} and \ref{f1b}) is in general not the same as in the total susceptibility (Eq.~\ref{f1c}). In the following, we shall thus limit our theoretical analysis to a calculation of the anisotropy of the susceptibility.

We note that $\chi^{\mathrm{mag}}_a > \chi^{\mathrm{mag}}_b$ was interpreted in Ref.~\onlinecite{Mingquan2017} as evidence of short-range magnetic order with the spins aligned along the shorter orthorhombic axis in FeSe. However, this explanation was found to be inconsistent with the absence of significant $^{57}$Fe NMR line broadening at low temperature~\cite{Tao2020}.

%-------------------------------------------------------------------------------------
\subsubsection{Knight-shift anisotropy: microscopic theory}
\label{Sec:theoryKS}
%-------------------------------------------------------------------------------------

\begin{figure*}[t!]
\includegraphics[width=0.7\textwidth]{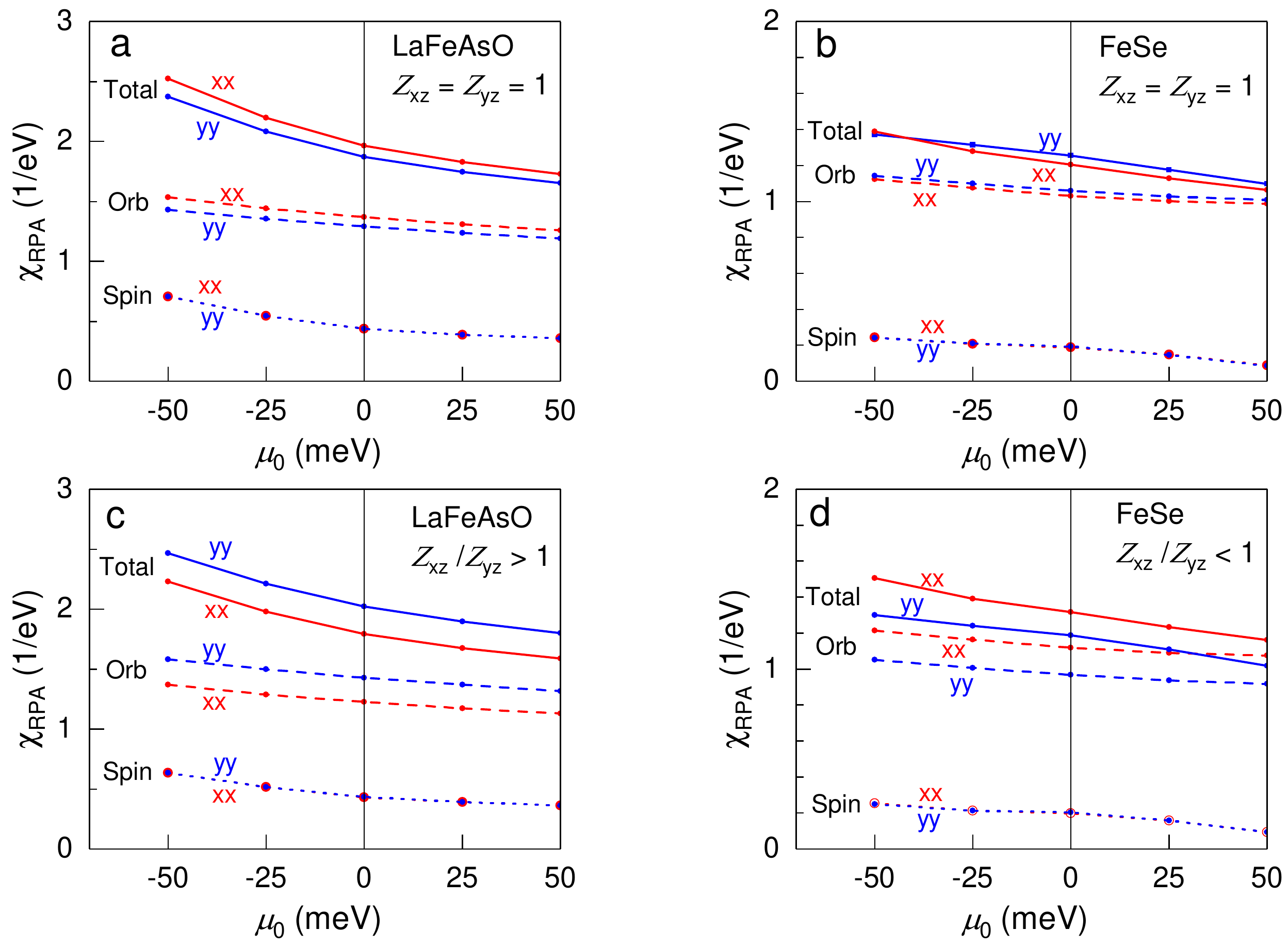}

%-----------------------------------------------------------
\caption{{\bf Calculated magnetic susceptibility with and without self-energy effects.} RPA susceptibilities ($U=4$eV) for the static limit of the uniform susceptibility as a function of chemical potential $\mu_{0}$, for the band structures of LaFeAsO (a,c) and FeSe (b,d), respectively. Solid, dashed, and dotted lines refer to total, orbital, and spin contributions to the uniform susceptibility, respectively. Blue and red lines refer to $\chi_{xx}(\vs{0},0)$ and  $\chi_{yy}(\vs{0},0)$, respectively. Panels (a,b) [(c,d)] are obtained without [with] additional self-energy effects, as described in the main text.}
\label{fig:theory_1}
\end{figure*}

The magnetic anisotropy of the spin fluctuations at $\vs{Q}_{1}$ and $\vs{Q}_{2}$ arising from SOC in both the normal and superconducting states was recently studied theoretically by two of the present authors within an itinerant ten-band model with multiorbital Hubbard interactions~\cite{scherer18,scherer19}. Here, we apply the theoretical framework developed in Ref. \onlinecite{scherer18} and extend it by including the paramagnetic orbital contribution to the susceptibility in order to obtain the total magnetic response throughout momentum space, and to extract the corresponding magnetic anisotropy at $ \vs{q} = (0,0)$, as well as at $\vs{Q}_{1}$ and $\vs{Q}_{2}$. For the bandstructure part of the Hamiltonian we apply DFT-derived models relevant to LaFeAsO and FeSe\cite{ikeda10,scherer17}. In the nematic phase, the LaFeAsO (FeSe) band contains a term describing ferro-orbital onsite (bond-orbital) nematic order, as suggested by previous studies\cite{Huh,Ding2015,Watson2016,Kim2013,kovacic14,Mukherjee2015,kreisel17,Sprau2017,scherer17,Kostin2018,Xingye2018}. For all model details, including the calculation of the full susceptibility tensor $\chi_{ij}(\vs{q},\omega)$ containing both spin and orbital contributions, we refer the reader to the SM section. Note we work with a coordinate system where $ x = a, y =  b, z =  c$.

In order to discuss the various contributions to the magnetic anisotropy, the susceptibility tensor is decomposed into three contributions of different physical origin
\begin{equation}
\chi_{ij}(\vs{q},\omega) = \chi_{ij}^{\mathrm{orb}}(\vs{q},\omega) + \chi_{ij}^{\mathrm{spin}}(\vs{q},\omega) + \chi_{ij}^{\mathrm{mixed}}(\vs{q},\omega).
\end{equation}
Here, $ \chi_{ij}^{\mathrm{orb}}(\vs{q},\omega) $ denotes the orbital part due to spin-conserving particle-hole fluctuations in the Fe $3d$ shell, $ \chi_{ij}^{\mathrm{spin}}(\vs{q},\omega) $ denotes the spin part, while $ \chi_{ij}^{\mathrm{mixed}}(\vs{q},\omega) $ contains the contributions that describe the response of angular momentum to changes in the electronic spin and vice versa. In the absence of SOC, this contribution vanishes identically, and with SOC being a small energy scale, the influence of $ \chi_{ij}^{\mathrm{mixed}}(\vs{q},\omega) $ on the total magnetic response and the magnetic anisotropy in particular, is negligible. In what follows, we will therefore focus the discussion on the orbital and spin contributions to the susceptibility, but all numerical results are obtained from the total $\chi_{ij}(\vs{q},\omega)$. The Knight-shift anisotropy is obtained from the real part of the static uniform susceptibility.

Using this theoretical framework, we explore the contributions to the total susceptibility and the roles played by nematicity and SOC in the generation of magnetic anisotropy. Without SOC or nematicity, $ \chi_{xx}^{\mathrm{orb}}(\vs{q},\omega) = \chi_{yy}^{\mathrm{orb}}(\vs{q},\omega) \neq  \chi_{zz}^{\mathrm{orb}}(\vs{q},\omega) $ due to the breaking of orbital rotational symmetry by hopping and crystal field terms, while the spin part remain fully symmetric. However, in the presence of both symmetry-breaking effects, SOC and nematicity, explicit calculations of $\chi_{ij}(\vs{q},\omega)$ (with interactions included at the RPA - random phase approximation - level, see SM) reveal that:
\begin{itemize}
\item{The magnetic anisotropy is momentum-selective in the sense that the orbital contribution dominates at $\vs{q}=\vs{0}$ while the spin contribution dominates at $\vs{Q}_1$ and $\vs{Q}_2$: $ \chi_{ij}(\vs{0},0) \approx \chi_{ij}^{\mathrm{orb}}(\vs{0},0) $ and $ \chi_{ij}(\vs{Q}_{1/2},0) \approx \chi_{ij}^{\mathrm{spin}}(\vs{Q}_{1/2},0) $. }
\item{The band structures and nematic orders relevant for LaFeAsO and FeSe, respectively, lead to the opposite magnetic Knight-shift anisotropy as compared to experiments, i.e. opposite inequalities in the expressions (\ref{expanisotropyLaFeAsO}) and (\ref{chi_FeSe}).}
\end{itemize}

The first point above, i.e. that the uniform susceptibility anisotropy is dominated by orbital contributions, is shown in Fig.~\ref{fig:theory_1}(a,b) and constitutes the first main outcome of our theoretical analysis. In Fig.~\ref{fig:theory_1} we plot the susceptibilities as a function of the chemical potential $\mu_{0}$ in a range around $\mu_{0} = 0$, where an electronic filling of $ n \approx 6 $ is realized. The results are plotted versus  $\mu_{0}$ in order to probe the sensitivity of the obtained results with respect to Fermi surface changes. As seen from Fig.~\ref{fig:theory_1}(a,b), the orbital anisotropy is largely insensitive to changes in the chemical potential.

The second point above, i.e. the opposite anisotropy compared to experiments, see Fig.~\ref{fig:theory_1}(a,b), exerts a serious problem for the "plain vanilla" RPA approach which cannot be alleviated by a small change of parameters, see SM. However, certainly for FeSe it is well-known that additional electronic interaction effects are required to explain, for example, the energy and momentum dependence of the magnetic fluctuations, or the superconducting pairing kernel\cite{kreisel17,Sprau2017,Kostin2018,Kreisel2018,martiny2019}. Indeed FeSe has been advocated to be an example of a Hund's metal where sizable orbital decoupling takes place as a consequence of large Hubbard-Hund interactions\cite{Yin2011,Medici_review,Roekeghem_review,Kostin2018}.

To this end, following earlier works\cite{kreisel17,Sprau2017,Kostin2018,Kreisel2018} we augment the RPA framework by phenomenological, orbital-selective quasiparticle weights $Z_\mu$ (here $ \mu $ is an orbital index, see SM) which turn out to play a crucial role in correctly determining the final splitting of orbital fluctuations with polarization along the crystal axes $ a $ and $ b $. The splitting is essentially controlled by the ratio $ Z_{xz} / Z_{yz} $ in the nematic phase. Methods capable of computing $ Z_{\mu} $ due to local correlations in the nematic phase predict that $ Z_{xz} / Z_{yz} > 1 $ for ferro-orbital onsite nematic order, while $ Z_{xz} / Z_{yz} < 1 $ for a mixture of $s-$ and $d-$wave bond-order~\cite{Mary,RongSi2018}. As elaborated in the SM section, we find that it is precisely this ratio $ Z_{xz} / Z_{yz} $ which controls the way the low-energy orbital fluctuations at small wavevectors split, and which further differentiates the magnetic anisotropy of LaFeAsO from that of FeSe. In Fig.~\ref{fig:theory_1}(c,d), we show the final total susceptibility with inverted anisotropy as compared to the "coherent" ($Z_\mu=1$) case displayed in Fig.~\ref{fig:theory_1}(a,b). Thus we conclude that for the in-plane Knight-shift anisotropy:
\begin{itemize}
\item{The experimental Knight-shift anisotropy for LaFeAsO and FeSe is theoretically reproduced by including self-energy corrections to the susceptibility in the form of orbital-dependent quasiparticle weights.}
\end{itemize}

We note that SOC does not play a role in the determination of the magnetic anisotropy due to orbital fluctuations. It is caused by the nematicity, which feeds back on the quasiparticle weights, and produces inverse ratios for  $ Z_{xz} / Z_{yz} $ for LaFeAsO and FeSe, respectively, resulting in final qualitative agreement with the experimental measurements.

We end this section with additional comments on the connection to experiments:

1) In the above scenario, quantitative agreement to the measured ratio $R_K=K_b/K_a$ can be reproduced by tuning $\delta Z$.  For example, in the case of FeSe, as seen from Fig.~\ref{fig:theory_1}(d), $R_K$ as determined simply from $\chi_{yy}(\vs{0},0)/\chi_{xx}(\vs{0},0)$ is of the order of 0.9 (similar to experiments) for $\delta Z=-0.05$. The fact that $Z_{yz} > Z_{xz}$ for FeSe is consistent with quasiparticle weights used earlier to model data from various experimental probes\cite{kreisel17,Sprau2017,Kostin2018,Kreisel2018,Cercellier2019}. For the latter case, however, a quantitatively larger quasiparticle weight anisotropy was required for agreement with the data. Only microscopic calculations that properly incorporate interactions and their feedback on the low-energy electronic states can reliably calculate $\delta Z$, which is a project beyond the scope of the current paper. We elaborate further on the dependence of the magnetic susceptibility on $\delta Z$ in the SM section.

2) The above discussion shows that it would be extremely interesting to have independent and more direct measurements of the quasiparticle weight anisotropy in iron pnictides.

3) Both nematicity and $\delta Z$ vanish at the nematic transition temperature, and therefore $R_K$ necessarily vanishes at that temperature as well. Below the nematic transition temperature, the detailed temperature dependence of the Knight-shift anisotropy will depend on the form of the $T$-dependence of both the nematic order and $\delta Z$. A standard mean-field like dependence would be consistent with experiments.

4) The magnetic anisotropy at $\vs{Q}_{1}$ and $\vs{Q}_{2}$ has been extensively explored by polarized neutron scattering experiments. In the SM section, we elaborate on the theoretical results for the anisotropy of  $\chi_{ij}(\vs{Q}_{1/2},0)$ and show its agreement with the available experimental data\cite{scherer18,Dai2015}

\subsubsection{$T_1$ anisotropy: experimental results}
\label{Sec:T1-results}

For both LaFeAsO and FeSe, the spin-lattice relaxation rate $1/T_1$, measured for each peak ({\it i.e.} for $H\parallel a$ and $H\parallel b$), becomes increasingly anisotropic upon cooling in the nematic state as seen from Fig.~\ref{asy}(b). While this agrees with earlier studies of both compounds~\cite{Baek2015,Fu2012}, our unambiguous site assignment allows us to discover that the $T_1$-anisotropy in FeSe ($R_{ab} \simeq 0.6$ at 10~K) is actually inverted with respect to that in LaFeAsO ($R_{ab} \simeq 2$ at $T_N$). This is also to be compared to $R_{ab} \simeq 1.6$ in NaFeAs~\cite{Zhou2016}.

To understand this result, it is first important to realize that $R_{ab}$ is {\it not} a measure of the $a$-$b$ anisotropy of spin fluctuations, that is, $R_{ab}\neq \chi^{\prime\prime}_b/\chi^{\prime\prime}_a$. This is because $1/T_1$ measures hyperfine field fluctuations transverse to the external field and because the hyperfine coupling tensor is non-diagonal at $^{75}$As and $^{77}$Se sites in Fe-based superconductors~\cite{Kitagawa2008}.

In fact, $R_{ab}$ is related to both the anisotropy of $\chi^{\prime\prime}$ (so-called spin-space anisotropy) and the $q$-space structure of low-energy spin fluctuations To realize this, it is first useful to consider fluctuations only at wave vectors $\vs{Q}_{1}=\left( \pi ,0 \right)$ and $\vs{Q}_{2}=\left( 0,\pi  \right)$ and an infinite correlation length $\xi$. In this limit, one obtains (see SM for details of the calculation and Ref.~\onlinecite{Kissikov2017} where identical formulas were recently derived):

\begin{equation}
\label{f3}
R_{ab}=\frac{\chi_a^{\prime \prime}(\vs{Q}_{1})+\chi_b^{\prime \prime}(\vs{Q}_{2})+\chi_c^{\prime \prime}(\vs{Q}_{1})}{\chi_a^{\prime \prime}(\vs{Q}_{1})+\chi_b^{\prime \prime}(\vs{Q}_{2})+\chi_c^{\prime \prime}(\vs{Q}_{2})}
\end{equation}
Above $T_s$, fluctuations at $\vs{Q}_1$ and $\vs{Q}_2$ have equal weight and thus $R_{ab}= 1$ , as indeed observed for LaFeAsO and NaFeAs. In the nematic state of this pnictides, spectral weight is progressively transferred from $\vs{Q}_2$ to $\vs{Q}_1$ upon cooling~\cite{Lu2014} and $R_{ab} \simeq 1+\frac{\chi_c^{\prime \prime}(\vs{Q}_{1})}{\chi_a^{\prime \prime}(\vs{Q}_{1})}$ exceeds 1 above $T_N$.

The less-than-unity and nearly inverse ratio $R_{ab}\simeq 0.6$ in FeSe could be understood in this picture if the dominant spin fluctuations are at $\vs{Q}_2$, not at $\vs{Q}_1$ as in the pnictides, thus leading to $R_{ab}\simeq1-\frac{\chi_c^{\prime\prime}(\vs{Q}_{2})}{\chi_b^{\prime\prime}(\vs{Q}_{2})+\chi_c^{\prime\prime}(\vs{Q}_{2})}$. This would imply that the magnetoelastic coupling in FeSe has an opposite sign as compared pnictides: the lattice would distort in an opposite way so that spin correlations are ferromagnetic (FM) along the $a$ axis and AFM along the short $b$ axis. There are, however,  problems with this explanation. First, it has been argued that the magneto-elastic coupling in FeTe, that is isostructural to FeSe, has the same sign as in pnictides~\cite{Paul11}, thus making a sign change in FeSe implausible. Second, a recent neutron scattering study on detwinned FeSe finds spectral weight at $\vs{Q}_1$. Third, for not too large values of the correlation length (a plausible situation given the absence of spin order in FeSe), Eq.~(\ref{f3}) is an oversimplification and other parameters play a role in the $T_1$ anisotropy: the value of $\xi$, its anisotropy as well as putative spectral weight away from $\vs{Q}_2$ and $\vs{Q}_1$.

Due to numerical limitations, we cannot obtain this quantity from microscopic calculations along the same lines as those presented above for the Knight shift. As seen from Eq.~(\ref{fs1}), a calculation of the $T_1$ anisotropy requires a reliable summation of all momenta in the BZ, which is very computationally demanding for ten-band models. Therefore, to understand the in-plane $T_1$ anisotropy, we shall pursue a phenomenological approach.

%-------------------------------------------------------------------------------------
\subsubsection{$T_1$ anisotropy: phenomenological calculation}
\label{Sec:theoryT1}
%-------------------------------------------------------------------------------------.

Assuming a functional form (Lorentzian here) of $\chi^{\prime\prime}(q_x,q_y)$ and knowing the $q$ dependence of the hyperfine form factor, the anisotropy ratio $R_{ab}$ can be calculated from Eq.~(\ref{fs1}) after integration over in-plane wave vectors $\vs{Q}=(q_x,q_y)$ (see SM for details). In addition, in order to further constrain the parameters, we have performed similar calculations for $R_{ac}=\frac{1}{2}((1/T_1)_a+(1/T_1)_b))/(1/T_1)_c$, the value of which has been measured in FeSe: $R_{ac}\simeq1.9\pm0.2$~\cite{Boehmer2015,Baek2015,Weiqiang2016}.

Perhaps surprisingly, there is still significant uncertainty regarding $\chi^{\prime\prime}(q,\omega)$ in FeSe: the value of the instantaneous spin-spin correlation length $\xi$ is essentially unknown and the low-energy fluctuations are poorly characterized. For instance, a recent inelastic neutron scattering (INS) study in the nematic state suggests that fluctuations at $\vs{Q}_2$ persist at energies of $\sim$3~meV and below, while they have totally disappeared in the range 6 to 11~meV~\cite{Tong2019}.

Given these uncertainties, we have performed extensive calculations of $R_{ab}$ and $R_{ac}$, varying three main parameters: (i) the peak position of $\chi^{\prime\prime}$ in the $(q_x,q_y)$ plane ({\it i.e.} we search whether low-energy fluctuations away from $\vs{Q}_1$ and $\vs{Q}_2$ may contribute to $T_1$), (ii) the correlation lengths $\xi_a$ and $\xi_b$ (defined as the inverse width of $\chi^{\prime \prime}$ along $q_x$ and $q_y$ directions) and (iii) the spatial anisotropy of $\chi^{\prime\prime}$.

Our results (detailed in SM) show that the experimental values of $R_{ab}$ and $R_{ac}$ cannot be simultaneously reproduced by assuming "standard" fluctuations at $\vs{Q}_1=(\pi,0)$, isotropic correlation length $\xi_a\simeq \xi_b$ (consistent with the isotropic scattering around $(\pi,0)$ in INS data at 15~meV~\cite{Zhao2016}), $\xi_{ab} \simeq 5 a_0$ (as might be expected, within an approximate factor of two, for a correlated material that does not order), $\chi^{\prime\prime}_c \simeq 7 \chi^{\prime\prime}_{b}$ (as indicated by INS data in the range of 2.5 to 8 meV~\cite{Mingwei2016}). Note that uncertainty regarding off-diagonal components of the hyperfine tensor in FeSe might quantitatively, but probably not qualitatively, affect the simulation results (see discussion in SM). First principes calculations of the hyperfine tensors would be helpful in order to progress on this issue.

As summarized in SM, correct $R_{ab}$ and $R_{ac}$ values are reproduced if one of the two following conditions is met:

\begin{itemize}
\item either the dominant spin fluctuations are peaked at $\vs{Q}_2$ and have a relatively isotropic but substantial correlation length: $\xi_a \simeq \xi_b \gtrsim$10$a_0$ (solution (3) in SM). In that case a relatively modest anisotropy of $\chi^{\prime\prime}$ in the $bc$ plane is required ($\chi^{\prime\prime}_b\simeq 1.5 \chi^{\prime\prime}_{c}$) while no constraint can be placed on $\chi^{\prime\prime}_a$.

\item or there is a large in-plane anisotropy of both $\xi$ (typically a factor 5) and $\chi^{\prime\prime}$ (typically a factor 10). Such large anisotropies  would be surprising in FeSe that does not order magnetically. An apparent anisotropy of $\xi$ could actually stem from the presence of competing spin-fluctuations at wave-vectors such as $(\pi,\pi/2)$ and $(\pi,\pi/3)$~\cite{Glasbrenner2015}. While anisotropic spectral weight has indeed been reported in Ref.~\onlinecite{Rahn2015}, this is not seen at low energy~\cite{Zhao2016} and so should be irrelevant for the low-energy fluctuations probed in NMR.

\end{itemize}

%-------------------------------------------------------------------------------------
\subsubsection{Discussion}
\label{Sec:discussion}
%-------------------------------------------------------------------------------------

We have showed that nematic order induces an in-plane anisotropy of both the Knight shift and the spin-lattice relaxation rate $1/T_1$ (i.e. a difference in the values measured for $H\parallel a$ and $H \parallel b$) that is reversed in FeSe with respect to LaFeAsO. The Knight shift results are consistent with earlier magnetization measurements comparing the in-plane magnetic anisotropy of nematic FeSe to the SDW phase of BaFe$_2$As$_2$\cite{Mingquan2017}.

Theoretically, starting from an itinerant scenario and realistic band structures, we calculated the static magnetic susceptibility $\chi^{\rm mag}$ including the effects of SOC and nematicity. The first important finding of our analysis is the dominance of the orbital part at $\vs{q}=\vs{0}$, presumably caused by the large number of orbital degrees of freedom per unit cell. We note that recent NMR experiments on FeSe find a dominant ($\geq 80\%$) orbital contribution to the Knight shift at low temperature, in agreement with our model.\cite{rxcao2018,Tao2020}  Also, the absence of any change in $\Delta K$ across the superconducting transition is consistent with an anisotropy arising from $K^{\rm orb}$ and a very small $K^{\rm spin}$ at low $T$ (see Fig.~\ref{asy}a where our lowest data point at 1.5 K should be well below $T_c$ even for 15~T applied parallel to the planes~\cite{Audouard}). By contrast, at  $\vs{Q}_{1}$ and $\vs{Q}_{2}$ the spin part strongly dominates, as shown in the SM section.

Then, we have found that the correct in-plane anisotropy of $\chi^{\rm mag}$ for FeSe and LaFeAsO can be reproduced only by postulating  orbital-dependent quasiparticle weights $Z_\mu$ such that $Z_{xz}/Z_{yz}<1$ for FeSe whereas $Z_{xz}/Z_{yz}>1$ for LaFeAsO. This difference is justified by the different types of nematic orders used to model the two materials: predominantly bond ordered in FeSe and predominantly ferro-orbital in LaFeAsO~\cite{Mary,RongSi2018}. These results quantitatively account for our Knight shift data.

At this point it is important to note that, particularly for FeSe, the low-energy electronic structure is still under intense investigation. For example, there is not yet consensus about the Fermi surface of the detwinned material~\cite{MattAmalia2018,Watson17,MYi,Huh,Rhodes20}, and e.g. the $T$-dependence of the low-energy bands and their orbital contents are under current scrutiny~\cite{MHC_trans,XLong}. In addition, the origin of nematicity in FeSe is still unsettled, and it remains unknown how the different Fe 3d orbital states participate in the nematicity. Proposals for distinct nematic orders in LaFeAsO versus FeSe include the itinerant spin-driven scenario for LaFeAsO~\cite{fernandes12,christensen16} whereas an instability caused by longer range Coulomb interactions in FeSe would naturally explain its mainly bond-ordered nature~\cite{jiang16,scherer17}.

From the above analysis we are led to the interesting conclusion that the opposite magnetic anisotropy between the nematic phases of FeSe versus LaFeAsO is caused by distinct nematicity in these two compounds, and its opposite feedback effects on the self-energy components for mainly the $d_{xz}$ and $d_{yz}$ orbitals. However, further support for this proposed scenario for iron-based superconductors requires a resolution to the origin of nematic order and its detailed orbital composition. In addition, we need a microscopic theoretical framework that self-consistently includes nematicity and self-energy effects.

Understanding the $T_1$ anisotropy is more involved. We found sets of parameters compatible with our experimental results (see Supplementary Information). However, the required strong anisotropies or the predominance of fluctuations at $\vs{Q}_2$ are difficult to reconcile with existing neutron scattering data. One could argue that the anomalous relaxation arises, not from spin fluctuations, but from a different type of magnetism such as orbital currents or from orbital fluctuations. However, this is unlikely as there is no experimental evidence of the former in any Fe-based material and the latter should be quenched deep in the nematic state. In fact, since the full parameter space of our phenomenological model could not be explored, it is possible that a solution exists with not too strongly anisotropic parameters and low-energy spin correlations having similar strength at $\vs{Q}_2$ at $\vs{Q}_1$.

In this context, it is interesting to notice that, according to a recent neutron scattering study, fluctuations at $\vs{Q}_2$ appear to have an anomalously high spectral weight at energies below $\sim$3~meV~\cite{Tong2019}. Therefore, we propose that the low-energy ($\sim\mu$eV) magnetic response of FeSe may not be simply deduced from the response at $\sim$10~meV and that the discrepancy is likely to arise from an enhanced weight of low energy fluctuations at $\vs{Q}_2$. It would therefore be extremely useful to further characterize the low-energy spin sector and to seek for theoretical explanations of low-energy fluctuations that are $C_4$ symmetric or nearly so.

%-------------------------------------------------------------------------------------
\subsubsection{Methods}
%-------------------------------------------------------------------------------------

Single-crystalline LaFeAsO and FeSe were synthesized by self-flux  and chemical vapor transport methods, respectively~\cite{SYan2009,SToulemonde2015}. The LaFeAsO single crystal was cut into a rectangle shape with its edges along the [100]$_{\rm o }$ and [010]$_{\rm o }$ of the orthorhombic (o) cell ({\it i.e.} Fe-Fe bond directions). The crystals were polished in order to obtain flat surfaces. The edges of the FeSe single crystal were naturally parallel to [100]$_{\rm o }$ and [010]$_{\rm o }$. The typical size of both samples is $1.5\times 1\times 0.1\text{m}{{\text{m}}^{\text{3}}}$.

The exact value of the external magnetic field was calibrated using the NMR line of metallic $^{63}$Cu (from the coil around the sample).

The uniaxial stress device (Fig.~\ref{unispectrum}e) is implanted on a semi-cylindrical Torlon holder, fitted in a cylindrical goniometer. Strain is applied through a BeCu sheet, and a pressure of about 10-20~MPa is obtained by tightening the screw by a quarter to half of a turn. By rotating the whole device, NMR spectra could be measured with the field carefully aligned either along the $b$ axis (parallel to the direction of applied strain) or along $a$ (perpendicular to the direction of applied strain). Line splittings were checked to vanish for $H$ tilted at 45$^\circ$ from $a$ and $b$.

The spin-lattice relaxation rate $T_1$ was measured by the saturation-recovery method for both LaFeAsO and FeSe. The time-dependence of the signal was fit using appropriate formulas for magnetic relaxation at (1/2 -1/2) transitions of nuclear spins 3/2 ($^{75}$As) and 1/2 ($^{77}$Se) with a single component at all temperatures ({\it i.e.} no stretching exponent).

The $^{75}$As Knight shift anisotropy of LaFeAsO was deduced after subtraction of the quadruolar shift (see supplementary materials).

%-------------------------------------------------------------------------------------
\subsubsection{Acknowledgments}
%-------------------------------------------------------------------------------------

We thank S. Karlsson and P. Strobel for their help in the growth of FeSe crystals as well as A. Boothroyd, V. Brouet, L. Fanfarillo, R. Fernandes, Y. Gallais, P.J. Hirschfeld, D. LeBoeuf, I. Mazin, C. Meingast, I. Paul, J. Schmalian, Y. Sidis, I. Vinograd, S. Wu, T. Wu, and particularly Igor Mazin, for discussions. Work in Grenoble was supported the Laboratoire d'excellence LANEF in Grenoble (ANR-10-LABX-51-01). Part of this work was performed at the LNCMI, a member of the European Magnetic Field Laboratory (EMFL). DDS and BMA acknowledge financial support from the Carlsberg Foundation.

%-------------------------------------------------------------------------------------

%-------------------------------------------------------------------------------------


\begin{references}

\bibitem{ARCMP}Fradkin, E.,  Kivelson, S. A.,  Lawler, M. J.,  Eisenstein, J. P. \&  Mackenzie, A. P. \emph{Annu. Rev. Condens. Matter Phys.} \textbf{1}, 153-178 (2010).

\bibitem{SNature98} Kivelson, S. A.,  Fradkin, E. \&  Emery, V. J. \emph{Nature} \textbf{393}, 550-553 (1998).

\bibitem{Yamase13}Yamase, H.  \&  Zeyher, R. \emph{Phys. Rev. B} \textbf{88}, 180502 (2013).

\bibitem{Maier14}Maier, T.  A.  \&  Scalapino, D.  J. \emph{Phys.  Rev.  B} \textbf{90}, 174510 (2014).

\bibitem{Lederer15}Lederer, S.,  Schattner, Y.,   Berg, E. \&  Kivelson, S. A. \emph{Phys. Rev. Lett.} \textbf{114}, 097001 (2015).

\bibitem{KNature15} Keimer, B.,  Kivelson, S. A., Norman, M. R., Uchida,  S. \&  Zaanen, J. \emph{Nature} \textbf{518}, 179-186 (2015).

\bibitem{Chu2010} Chu, J.-H. et al. \emph{Science} \textbf{329}, 824-826 (2010).

\bibitem{Fernandes2014}Fernandes, R. M.,  Chubukov, A. V. \&  Schmalian, J. \emph{Nat. Phys.} \textbf{10}, 97-104 (2014).

\bibitem{Dai2015}Dai, P. \emph{Rev. Mod. Phys.} \textbf{87}, 855 (2015).

\bibitem{Bascones2016}Bascones,  E., Valenzuela, B. \& Calderon, M. J.  \emph{C. R.  Physique} \textbf{17}, 36-59 (2016).

\bibitem{Cava2009}McQueen, T. M. et al.  \emph{Phys. Rev. Lett.} \textbf{103}, 057002 (2009).

\bibitem{Boehmer2015}B\"{o}hmer, A. E. et al.  \emph{Phys. Rev. Lett. }\textbf{114}, 027001 (2015).

\bibitem{Baek2015}Baek,  S.-H. et al. \emph{Nat. Mater.} \textbf{14}, 210-214 (2015).

\bibitem{Yamakawa2016}Yamakawa, Y., Onari, S. \& Kontani, H. \emph{Phys. Rev. X}  \textbf{6}, 021032 (2016).

\bibitem{Rahn2015}Rahn, M. C., Ewings,  R. A., Sedlmaier,  S. J., Clarke,  S. J. \& Boothroyd, A. T.  \emph{Phys. Rev. B} \textbf{91}, 180501(R) (2015).

\bibitem{ZhaoMat2016}Wang, Q. et al. \emph{Nat. Mat.} \textbf{15}, 159-163 (2016).

\bibitem{Mingwei2016}Ma, M. et al.  \emph{Phys. Rev. X } \textbf{7}, 021025 (2017).

\bibitem{Zhao2016}Wang, Q. et al.  \emph{Nat. Commun.} \textbf{7}, 12182 (2016).

\bibitem{Tong2019}Chen, T. et al. \emph{Nature Mat.} \textbf{18}, 709  (2019).

\bibitem{Schmalian2015}Chubukov, A. V.,  Fernandes, R. M. \& Schmalian, J. \emph{Phys. Rev. B} \textbf{91}, 201105 (2015).

\bibitem{FeSe1}Liebsch, A. \& Ishida, H.  \emph{Phys Rev. B} \textbf{82}, 155106 (2010). %,

\bibitem{FeSe2}Essenberger,  F.,  Buczek, P., Ernst,  A.,  Sandratskii, L. \&  Gross E. K. U.  \emph{Phys. Rev. B} \textbf{86}, 060412(R) (2012). %

\bibitem{FeSe3}Hirayama, M.,  Misawa,  T., Miyake, T. \&  Imada, M. \emph{J. Phys. Soc. Jpn.} \textbf{84}, 093703 (2015). %

\bibitem{FeSe4}Busemeyer,  B., Dagrada, M., Sorella, S., Casula, M. \&  Wagner, L. K. \emph{Phys. Rev. B} \textbf{94}, 035108 (2016). %

\bibitem{FeSe5}Liu, K., Lu, Z.-Y. \& Xiang, T.  \emph{Phys. Rev. B} \textbf{93}, 205154 (2016). %

\bibitem{FeSe6}Wang, Z., Hu, W.-J. \& Nevidomskyy, A. H.  \emph{Phys. Rev. Lett.} \textbf{116}, 247203 (2016). %

\bibitem{FeSe7}Lai, H.-H.,  Hu, W.-J., Nica, E. M., Yu, R. \& Si, Q. \emph{Phys. Rev. Lett.} \textbf{118}, 176401 (2017). %

\bibitem{Mukherjee2015}Mukherjee,  S., Kreisel, A., Hirschfeld, P. J. \&  Andersen, B. M.  \emph{Phys. Rev. Lett.} \textbf{115}, 026402 (2015).

\bibitem{Glasbrenner2015}Glasbrenner, J. K. et al.  \emph{Nat. Phys.} \textbf{11}, 953-958 (2015).

\bibitem{Lee2015}Wang, F., Kivelson,  S. A. \& Lee, D.-H. \emph{Nat. Phys.} \textbf{11}, 959-963 (2015).

\bibitem{scherer17}Scherer, D. D. et al. \emph{Phys. Rev. B} {\bf 95}, 094504 (2017).

\bibitem{Fanfarillo2018}Fanfarillo,  L., Benfatto, L. \&  Valenzuela, B. \emph{Phys. Rev. B} \textbf{97}, 121109(R) (2018).
%

\bibitem{Kreisel2018}Kreisel, A.,  Andersen, B. M. \&  Hirschfeld, P. J. \emph{Phys. Rev. B}  \textbf{98}, 214518 (2018).

\bibitem{AnneAndreas2017}B\"{o}hmer, A. E. \& Kreisel, A. \emph{J. of Phys.: Cond. Mat.} \textbf{30}, 023001 (2017).

\bibitem{MattAmalia2018}Coldea,  A. I. \& Watson, M. D.  Annu.  \emph{Rev. Condens. Matter Phys.} \textbf{9}, 125-146 (2018).

\bibitem{Sprau2017}Sprau, P.O. et al. \emph{Science} \textbf{357}, 75 (2017).

\bibitem{Kostin2018}Kostin, A. et al. \emph{Nature Mat.} \textbf{17}, 869 (2018).

\bibitem{Onari16}Onari, S.,  Yamakawa, Y. \& Kontani, H.  \emph{Phys. Rev. Lett.} {\bf 116}, 227001 (2016).

\bibitem{Nica17}Nica, E. M., Yu, R. \&  Si, Q. \emph{ npj Quantum Mater.} {\bf 2}, 24 (2017).

\bibitem{kreisel17}Kreisel, A.  et al. \emph{ Phys. Rev. B} {\bf 95}, 174504 (2017).

\bibitem{Benfatto18}Benfatto, L.,  Valenzuela, B. \& Fanfarillo, L. \emph{npj Quantum Mater.} {\bf 3}, 56 (2018).

\bibitem{Kang2018}Kang, J.,  Fernandes, R. M. \& Chubukov, A. \emph{Phys. Rev. Lett.} {\bf 120}, 267001 (2018).

\bibitem{She17}She,  J.-H.,  Lawler, M. J. \&  Kim, E.-A. \emph{Phys. Rev. Lett.} {\bf 121}, 237002 (2018).

\bibitem{Fanfarillo2016}Fanfarillo, L. et al. \emph{Phys. Rev. B} \textbf{94}, 155138 (2016).

\bibitem{Watson17}Watson,  M. D.,  Haghighirad, A. A.,  Rhodes, L. C., Hoesch,  M. \&  Kim, T. K. \emph{ New J. Phys.} \textbf{19}, 103021 (2017). %

\bibitem{MYi}Yi, M. et al. \emph{Phys. Rev. X} \textbf{9}, 041049 (2019).

\bibitem{Huh}Huh, S. et al. \emph{Commun Phys} \textbf{3}, 52 (2020).

\bibitem{Rhodes20}Rhodes,  L. C., Watson, M. D., Haghighirad, A. A., Evtushinsky, D. V. \&  Kim, Y. K. \emph{Phys. Rev. B} \textbf{101}, 235128 (2020).

\bibitem{XLong}Long,  X., Zhang,  S.,  Wang, F. \& Liu,  Z. \emph{npj Quantum Materials} \textbf{5}, 50 (2020).

\bibitem{MHC_trans}Christensen, M. H., Fernandes,  R. M. \&  Chubukov, A. V.  \emph{Phys. Rev. Research} \textbf{2}, 013015 (2020).

\bibitem{Lu2014}Lu, X. et al. \emph{Science} \textbf{345}, 657-660 (2014).

\bibitem{Christensen2015}Christensen, M. H., Kang, J., Andersen, B. M., Eremin, I. \&  Fernandes, R. M.  \emph{Phys. Rev. B} \textbf{92}, 214509 (2015).

\bibitem{scherer18}Scherer D. D. \& Andersen, B. M. \emph{Phys. Rev. Lett.} {\bf 121}, 037205 (2018).

\bibitem{Zhou2016}Zhou, R.,  Xing, L. Y., Wang,  X. C., Jin,  C. Q. \&  Zheng, G.-q. \emph{ Phys. Rev. B} \textbf{93}, 060502 (2016).

\bibitem{Fu2012}Fu, M. et al.  \emph{Phys. Rev. Lett.} \textbf{109}, 247001 (2012).

\bibitem{Kissikov2017}Kissikov, T. et al. \emph{ Nat. Commun.} \textbf{9}, 1058 (2018).

\bibitem{He2016}He, M. et al. \emph{Nat. Commun.} \textbf{8}, 504 (2017).

\bibitem{Cruz2008}de la Cruz, C. et al. \emph{Nature} \textbf{453}, 899-902 (2008).

\bibitem{rxcao2018}Cao, R. X. et al. \emph{New J. Phys.} \textbf{21}, 103033 (2019).

\bibitem{Tao2020}Li, J. et al.  \emph{Phys. Rev. X} \textbf{10}, 011034 (2020).

\bibitem{Ok2018}Ok, J. M. et al. \emph{Phys. Rev. B} \textbf{97}, 180405 (2018).

\bibitem{Kitagawa2008}Kitagawa, K., Katayama, N., Ohgushi, K., Yoshida, M. \& Takigawa, M. \emph{J. Phys. Soc. Jpn.} \textbf{77}, 114709 (2008).

\bibitem{Mingquan2017}He, M. et al. \emph{Phys. Rev. B} \textbf{97}, 104107 (2018).

\bibitem{scherer19}Scherer, D. D. \& Andersen, B. M. Preprint at https://arxiv.org/abs/1906.08566 (2019).

\bibitem{ikeda10}Ikeda, H., Arita, R., \&  Kune{\v s}, J.  \emph{Phys. Rev. B} {\bf 81}, 054502 (2010).
%

\bibitem{Ding2015}P. Zhang, et al. \emph{Phys. Rev. B} \textbf{91}, 214503 (2015).

\bibitem{Watson2016}Watson, M. D. et al. \emph{Phys. Rev. B} \textbf{94}, 201107 (2016).

\bibitem{Kim2013}Kim, Y. K.  et al. \emph{Phys. Rev. Lett.} \textbf{111}, 217001 (2013).

\bibitem{kovacic14}Kovacic, M., Christensen, M. H., Gastiasoro, M. N. \&  Andersen, B. M. \emph{Phys. Rev. B} {\bf 91}, 064424 (2015).

\bibitem{Xingye2018}Lu, X. et al.  \emph{Phys. Rev. Lett.} \textbf{121}, 067002 (2018).

\bibitem{martiny2019}Martiny, J. H. J., Kreisel, A. \& Andersen, B. M. \emph{Phys. Rev. B}  \textbf{99}, 014509 (2019).

\bibitem{Yin2011}Yin, Z. P., Haule, K. \&  Kotliar, G.  \emph{Nature Mat.} \textbf{10}, 932 (2011).

\bibitem{Medici_review}Medici, L \& Capone, M. in  Mancini, F. \& Citro,  R. (eds) \emph{The Iron Pnictide Superconductors :An Introduction and Overview.} (Springer Series in Solid-State Sciences 186 , Cham, 2017).

\bibitem{Roekeghem_review}Roekeghem. A., Richard, P., Ding, H. \& Biermann, S. \emph{ C. R. Physique} {\bf 17}, 140 (2016).

\bibitem{Mary}Chatzieleftheriou, M. \emph{M.Sc. thesis}, University of Copenhagen (2017).

\bibitem{RongSi2018}Yu, R.,  Zhu, J.-X. \& Si,  Q.  \emph{Phys. Rev. Lett.} \textbf{121}, 227003 (2018).

\bibitem{Cercellier2019}Cercellier, H., Rodi\`ere, P., Toulemonde, P., Marcenat, C. \& Klein, T.  \emph{Phys. Rev. B} \textbf{100}, 104516 (2019).
%

\bibitem{Paul11}Paul, I., Cano, A. \& Sengupta, K. \emph{Phys. Rev. B} \textbf{83}, 115109 (2011).

\bibitem{Weiqiang2016}Wang, P. S. et al. \emph{Phys. Rev. Lett.} \textbf{117}, 237001 (2016).

\bibitem{Audouard}Audouard, A. et al.  \emph{EPL} \textbf{109}, 27003 (2015).

\bibitem{fernandes12}Fernandes, R. M., Chubukov, A. V., Knolle, J., Eremin, I. \& Schmalian,  J.  \emph{Phys. Rev. B} \textbf{85}, 024534 (2012).

\bibitem{christensen16}Christensen, M H.,  Kang, J., Andersen,  B. M. \&  Fernandes, R. M. \emph{ Phys. Rev. B} \textbf{93}, 085136 (2016).

\bibitem{jiang16}Jiang, K., Hu,  J., Ding, H. \& Wang,  Z. \emph{Phys. Rev. B} \textbf{93}, 115138 (2016).

\bibitem{SToulemonde2015}Karlsson, S., et al. \emph{Supercond. Sci. Technol.} \textbf{28} 105009 (2015).

\bibitem{SYan2009}Yan, J.-Q. et al.  \emph{Appl. Phys. Lett.} \textbf{95}, 222504 (2009).



%----------------------------------------------

\end{references}
\end{document}